\documentclass[a4paper]{jpconf}
\usepackage{iopams,cite}
\usepackage{graphicx}
\usepackage{amsfonts}
\usepackage{mathrsfs}
\usepackage{epstopdf}
\usepackage{bm}

\newcommand{\be}{\begin{equation}}
\newcommand{\ee}{\end{equation}}
\newcommand{\bea}{\begin{eqnarray}}
\newcommand{\eea}{\end{eqnarray}}

\newcommand{\ket}[1]{|#1\rangle}

\renewcommand{\Re}{{\rm Re}}
\renewcommand{\Im}{{\rm Im}}

\begin{document}

\title{Non-hermitian approach to decaying ultracold bosonic systems}

\author{Sandro Wimberger, Carlos A. Parra-Murillo, and Georgios Kordas}
\address{Institut f\"ur Theoretische Physik and Center for Quantum Dynamics, 
         Universit\"at Heidelberg, D--69120 Heidelberg}

\ead{s.wimberger@thphys.uni-heidelberg.de}

\begin{abstract}
A paradigm model of modern atom optics is studied, strongly interacting ultracold bosons in an optical lattice. This many-body system can be artificially opened
in a controlled manner by modern experimental techniques. We present results based on a non-hermitian effective Hamiltonian whose quantum spectrum is analyzed.
The direct access to the spectrum of the metastable many-body system allows us to easily identify relatively stable quantum states, corresponding to previously 
predicted solitonic many-body structures.
\end{abstract}

%\pacs{03.65.Yz}{Decoherence; open systems; quantum statistical methods}
%\pacs{03.75.Lm}{Tunneling, Josephson effect, Bose-Einstein condensates in periodic potentials, solitons, vortices, and topological excitations}
%\maketitle

\section{Introduction}
\label{intro}\medskip

Open quantum systems are ubiquitous in microscopic physics. Any system can only be decoupled from its environment in an abstract sense.
There exist many successful approaches to describe open quantum systems. Let us just mention the theory of scattering \cite{taylor}, techniques based on 
Master equations \cite{breuer}, heavily used e.g. in quantum optics \cite{WM}, or effective Hamiltonians which contain absorbing boundary conditions \cite{rotter}.
The latter approach has the advantage that it allows one to access a generalized quantum spectrum, thus it extends spectral methods, normally applied to closed systems,
to open ones. Keeping in mind that the method of effective non-hermitian Hamiltonians is based on perturbative arguments, such as a weak decay or coupling to the environment,
it nevertheless gives the possibility to predict the evolution of the system for short times at least. The qualitative understanding based on the analysis of the spectrum
 is at least as useful as more accurate methods which are based on numerical propagation in most cases. 

\begin{figure}[ht]
\includegraphics[width=18pc]{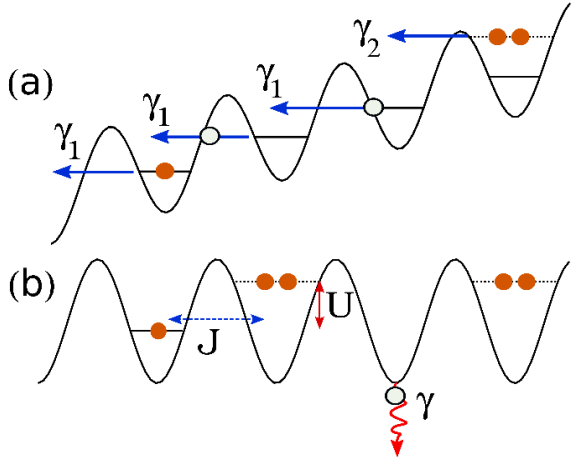}\hspace{2pc}
\begin{minipage}[b]{18pc}
\caption{\label{fig:1} 
Sketch of two typical realization of strongly correlated many-body open quantum systems: (a) Particles tunneling from a tilted optical lattice.
The local decay rates depend on the initial particle numbers in the wells represented by the indices, see reference \cite{TMW} for details.  
(b) engineered particle loss with decay rate $\gamma$ in one of the potential wells. The terms $J$ and $U$ denote the
hopping and the two-body interaction energy respectively. Interplay of the various processes can lead to interesting dynamics in the open many-body
systems, as discussed in section \ref{results} for case (b).
}
\end{minipage}
\end{figure}

In this paper we concentrate on systems of ultracold atoms loaded into periodic lattice structures. Such systems are naturally open when one, for instance, starts to tilt 
or to accelerate the lattice \cite{AW11} since this induces decay or quantum tunneling. 
Alternatively, the system may be opened artificially to observe interesting effects induced by 
dissipation and/or quantum noise. Figure \ref{fig:1} sketches the mentioned setups. While we studied the case (a) of tilted lattices for strongly correlated bosons in 
previous works \cite{TMW,AW11,Ploetz,carlos}, in the following we investigate the case (b) in more detail with an effective non-hermitian Hamiltonian for the many-body system. 
Experimentally, the loss of particles as shown in figure \ref{fig:1}(b) can be realized easily in a controlled manner. Single particle decay channels are provided by
shining in appropriate laser beams \cite{Sher10} or even electronic beams \cite{ott1,ott2}. 
The latter ionize single atoms in the condensate, which, together with the produced electrons, 
are afterward accelerated by electric fields, hence they leave the system extremely fast without strong backaction to the remaining bosons in the lattice environment.

\section{A dissipative Bose-Hubbard system}
\label{results}\medskip

\subsection{Motivation}
\label{moti}\medskip

Ultracold bosons at zero temperature can be described by the following Bose-Hubbard Hamiltonian \cite{BDZ}
\begin{equation}
  \hat H = - \frac{J}{2}  \sum \nolimits_{j} \left( \hat a_{j+1}^{\dagger} \hat a_j + \hat a_{j}^{\dagger} \hat a_{j+1} \right)
         + \frac{U}{2} \sum \nolimits_j \hat a_{j}^{\dagger}  \hat a_{j}^{\dagger} \hat a_{j}  \hat a_{j} \,,
    \label{eq:1}
\end{equation}
provided the optical lattice holding them is sufficiently deep (such that the approximation by discrete modes denoted by the
index $j$ is good). Here $\hat a_j$ and $\hat a_j^\dagger$ are the bosonic annihilation and creation operators in the $j$th well. 
$\hbar$ is set to one measuring all energies in frequency units. This Hamiltonian can be made dimensionless by rescaling with one of the two energy scales
$J$ or $U$ respectively. Hence, without loss of generality, we set $J/2=1$ in the following, measuring all energies in units of $J/2$.

As briefly mentioned in the introduction and motivated in a series of previous papers on dissipative Bose condensates \cite{WTW08,leaky1,leaky2,KWW12}, a tunable
source of single-particle dissipation can be added to the system. With the spatially local dissipation rates $\gamma_j$, the dynamics
is then best described by a quantum Master equation \cite{breuer,ang97} for the many-body density $\hat \rho$:
\begin{equation}
 \dot{\hat \rho} = -i [\hat H,\hat \rho]
                 - \frac{1}{2}  \sum \nolimits_{j}  \gamma_{j} \left( \hat a_j^\dagger \hat a_j \hat \rho + \hat \rho \hat a_j^\dagger \hat a_j  
                 - 2 \hat a_j \hat \rho \hat a_j^\dagger \right) \,.
\label{eq:2}
\end{equation}
This Master equation may be ``unraveled'', i.e. solved exactly, by a quantum jump approach \cite{breuer}. Alternatively, it can be solved approximately by methods which take into
account effects beyond the mean-field expansion, see, e.g. \cite{leaky1,leaky2,KWW12}, where results of both methods are shown and compared.

Here we will follow a different approach based on an effective Hamiltonian in which the dissipative terms of equation~(\ref{eq:2}) are included
as complex absorbing potentials. In contrast to the above mentioned methods for solving the Master equation, this gives us access to the quantum spectrum
of the non-hermitian Hamiltonian studied in the next subsection. Within this approximation, the real parts of the spectrum correspond to the energy levels --
just as for any closed quantum system -- and the imaginary parts describe the decay widths of the metastable eigenstates, i.e. their stability in the presence 
of the loss \cite{rotter}.

\subsection{A spectral approach}
\label{approach}\medskip

As a study case, we now look at a particular situation of three potential wells (also called sites) with relatively large filling factors, i.e. particles per site, 
of the order ten or larger. For this setup, reference \cite{leaky1} predicts the {\it dynamical} formation of very stable collective states of many bosons within 
a single well, despite strong atom losses. More precisely, we choose $\gamma_1=\gamma_3=\gamma$ at the two boundary wells and $\gamma_2=0$ in the middle site. 
For weak atomic interactions,
i.e. for the limit $U\to0$, all particles will be lost steadily as time evolves. This is just the situation for independent (almost) non-interacting particles. For strong
interactions, i.e. $U > U_{\rm crit}$, however, the particles at the leaky sites one and three are quickly lost, while the well in the middle is protected against the decay.
This may be expected by a simple mean-field argument, saying that the bosons cannot get rid of their high energy in the middle well by just gaining the energy $J=2$ by
hopping to the leaky sites. Then, $U_{\rm crit}\bar{n} \gtrsim J=2$ in our dimensionless units, 
which is the so-called self-trapping criterion where mean-field tunneling is suppressed by the mentioned energetic argument \cite{leggett}. 
$\bar{n}$ is the filling factor of the corresponding well. The big surprise reported in \cite{leaky1,leaky2} is yet that the surviving part of the initially
prepared condensate is very coherent, implying that non-trivial many-body dynamics occurred during the formation of this very stable, so-called breather or 
solitonic state. This strongly correlated state is well known for its stability with respect to decay \cite{Fla08}, and this property we are going to use in order to identify 
them using our spectral analysis. Thus dissipation can be used to prepare almost pure condensed many-body states in the middle well \cite{leaky1,leaky2}. 

Those breather states can be easily detected (without any time-propagation as done in \cite{leaky1,leaky2}), by studying the spectrum of the following effective Hamiltonian
\begin{equation}
  \hat H_{\rm eff} = - \sum_{j=1}^{3} \left( \hat a_{j+1}^{\dagger} \hat a_j + \hat a_{j}^{\dagger} \hat a_{j+1} 
         + \frac{U}{2} \hat a_{j}^{\dagger}  \hat a_{j}^{\dagger} \hat a_{j}  \hat a_{j} - i \frac{\gamma}{2} \hat{n}_j \right)\,,
    \label{eq:3}
\end{equation}
with $\hat{n}_j=\hat a_j^{\dagger} \hat a_j$ being the number operator. In the framework of the quantum jump method, this effective Hamiltonian
gives the continuous evolution between two quantum jumps. Nevertheless, one can use it to identify and study the breather state, which is predicted as
an attractively stable state of the evolution of the full open system. 
Here, for simplicity, we use periodic boundary conditions for $\hat H_{\rm eff}$, identifying the well
four by the well one. The results presented below qualitatively remain unchanged when using open (also called hard-wall) boundary conditions, since the decay
affects both ends of the lattice in exactly the same way.

\begin{figure}[ht]
\includegraphics[width=18pc]{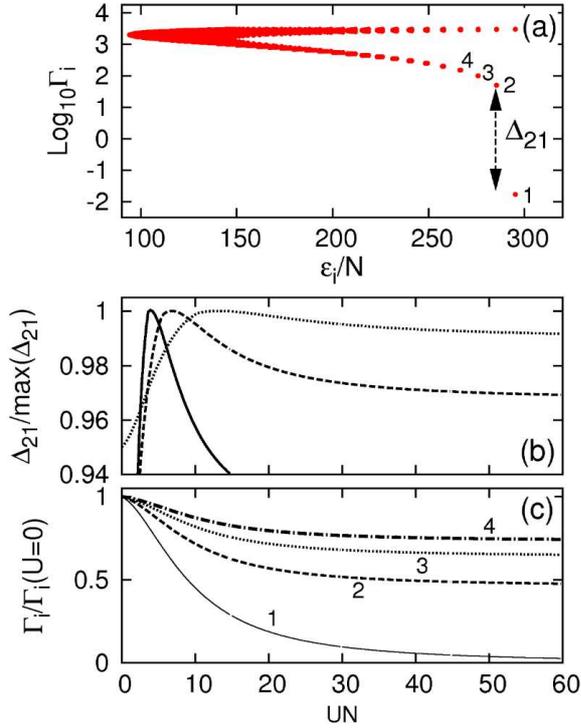}\hspace{2pc}
\begin{minipage}[b]{18pc}\caption{
(a) Spectrum of the effective Hamiltonian (\ref{eq:3}) for a loss rate $\gamma=20$, interaction strength $U=2$ and fixed number of particles $N=60$. 
Shown are $\Gamma_i=-2\Im{E_i}$ vs $\epsilon_i /N = \Re{E_i}/N$. (b) The spectral gap $\Delta_{21}$ from (a) as a function of $UN$ for $\gamma=5$ (solid line), 
$\gamma=10$ (dashed line), and $\gamma=20$ (dotted line). 
The values on the $y$-axis are normalized with respect to their maximum in order to allow a better comparison. (c) The decay rates of the four most stable
states vs $UN$ for $\gamma=20$. In (b,c) $N=60$ is always kept fixed and $U$ was varied. Doing the same for a different value of $N \gg 1$ produces similar results,
hence the results are robust as long as $N \gg 1$ (which is necessary for the scaling with $UN$ to be approximately valid).
\label{fig:2} }
\end{minipage}
\end{figure}

Figure \ref{fig:2} reports in panel (a) the eigenspectrum $\{E_k\}_k$ of the effective Hamiltonian $\hat H_{\rm eff}$ for typical parameters. There, the decay rates, i.e. 
$\Gamma_k=-2\Im{E_k}$, are plotted versus the real parts of the spectrum $\epsilon_k = \Re{E_k}$. 
One observes immediately one state which is far separated from the rest of the spectrum. This is the very stable breather state discussed above. Its stability is given by the 
rate $\Gamma_1$, where the index starts to count from the most stable state. 
As the effective interaction $U \bar{n}$ increases, this state moves away from the rest of the spectral band.
A criterion for the stability is therefore the distance of the imaginary parts from the next stable state (which actually is degenerate because of the spatial symmetry 
of the problem, and both are denoted by the index two in panel (a)). This distance, or spectral gap $\Delta_{21}=\Gamma_2-\Gamma_1$, 
is plotted in panel (b) as a function of the effective
interaction $UN$, where $N$ denotes the total particle number in the system (which is fixed to a specific value). We can indeed confirm that the breather state forms 
for a sufficiently large effective interaction strength. The critical value may be estimated by the maxima of the curves shown in panel (b). These maxima come about because
all the lower lying states in the spectrum have decreasing decay rates $\Gamma_k$ as a function of $UN$, yet the one which decreases fastest is exactly the breather state,
the one with minimal $\Gamma_1$ in panel (c). 

The qualitative picture remains unchanged for a large range of single-particle loss rates $\gamma$ in the effective Hamiltonian of equation~(\ref{eq:3}), as shown in 
figure~\ref{fig:2}(b). Increasing $\gamma$ tends to widen the peak in the spectral gap $\Delta_{21}=\Gamma_2-\Gamma_1$ and to shift the peak position toward larger values
of $UN$ (since more interaction energy is necessary to balance the increasing loss at the outer sites, i.e. to avoid decay into them).
An interesting effect can be observed for the very large dissipation rate $\gamma=20$. As one can see in figure~\ref{fig:2}(b), the 
normalized spectral gap is almost one even for zero interaction in this case. This means that the decay rate is almost independent of the interaction strength.
This happens because strong dissipation blocks the tunneling to the two leaky sites, an effect discussed in detail already in \cite{leaky1,leaky2}.

The most stable eigenstate of $\hat H_{\rm eff}$ with decay rate $\Gamma_1$ corresponds to almost one hundred percent to the Fock state with all particles in the middle site, 
represented by $\ket{0,N,0}$. This is expected since this state has the largest real part, given essentially by the on-site effective interaction $UN$, and hence it is most
stable by energetic constraints which forbid the hopping of single particles to the leaky boundary wells.

As a final remark, one may find analogies to the phenomenon of resonance trapping in open systems often described by effective non-hermitian Hamiltonians \cite{rotter,trapping}. 
This phenomenon 
describes the formation of a gap in the imaginary parts of the spectrum when the opening of the system is increased. The consequence is that very few very stable states
exist as compared with a large number of fast decaying ones. This effect appears as the opening, described by $\gamma$ in our case, increases, whilst in our problem
presented above, the stable breather state rather forms when increasing the interparticle interaction strength (and remaining qualitative unchanged over a wide range
of $\gamma$). Nevertheless it would certainly be interesting to further investigate this analogy for many-body quantum systems.

\section{Conclusions} 
\label{concl}\medskip

We have shown that the method based on effective Hamiltonians, as the one of equation~(\ref{eq:3}), is a useful tool for the quick identification of states with certain 
properties, in our case very stable many-body modes. 
A direct diagonalisation is always possible for small systems, i.e. not too many bosons and lattice sites. For larger many-body systems, a 
diagonalisation based on the Lanczos method may be used to find interesting states locally in the energy spectrum \cite{carlos}. For this, the scaling of eigenstates with the system size for 
smaller systems can guide the search for the position in the complex energy plane. In the future our results will be extended to larger systems, comparing with the
approximate evolution techniques used, for instance, in references \cite{leaky1,leaky2,KWW12} and extending results on one-band systems \cite{TMW} such as 
sketched in figure \ref{fig:1}(a) to several energy bands.

\ack
S W and G K thank with pleasure the organiser Thomas Elze for his kind invitation to DICE2012.
We are very grateful as well to Dirk Witthaut for our longstanding collaboration. Our work is supported by the 
the DFG (FOR760), the Helmholtz Alliance Program EMMI (HA-216), and the HGSFP (GSC 129/1).

\end{document}